\RequirePackage{fix-cm}

\documentclass[smallcondensed]{svjour3}     
\smartqed  
\usepackage{graphicx}
\usepackage[numbers, comma, compress, sectionbib]{natbib}
\usepackage{url}
\usepackage{setspace}
\usepackage[colorlinks,urlcolor=blue,citecolor=blue]{hyperref}
\usepackage{graphicx}
\usepackage{booktabs}
\usepackage{tabularx}
\usepackage{multirow}
\usepackage[utf8]{inputenc}
\usepackage[T1]{fontenc}
\usepackage[english]{babel}
\usepackage{enumitem}
\usepackage{color}
\usepackage{amsmath,amsfonts,amssymb}
\usepackage{url}
\usepackage{xspace}
\usepackage[noend]{algpseudocode}
\usepackage{algorithm}
\usepackage{algorithmicx}

\algrenewcommand\algorithmicrequire{\textbf{Input:}}
\algrenewcommand\algorithmicensure{\textbf{Output:}}

\usepackage{nicematrix}
\usepackage{tikz,pgfplots}
\pgfplotsset{compat=newest}
\usepackage{subcaption}
\usetikzlibrary{calc, chains,
                fit,
                positioning,
                shapes}
\usepgfplotslibrary{fillbetween}

\usetikzlibrary{automata,arrows,positioning,calc}
\definecolor{mygray}{RGB}{211,211,211}
\newcommand{\F}{\mathbb{F}_2}

\newcommand{\dgv}{\ensuremath{d_{GV}}}

\newcommand{\wt}{\mathrm{wt}}
\newcommand{\etal}{\emph{et al.}\xspace}

\newcommand{\lar}{\stackrel{\mathdollar}{\leftarrow}}
\newcommand{\sk}{\ensuremath{\mathsf{sk}}\xspace}
\newcommand{\pk}{\ensuremath{\mathsf{pk}}\xspace}
\newcommand{\WRH}{\textsf{WRH}}

\newcommand{\highlight}[1]{\textcolor{red}{#1}}
\renewcommand{\highlight}[1]{\textcolor{black}{#1}}

\begin{document}

\title{Cryptanalysis of a code-based full-time signature\thanks{This work was partially funded by the French DGA. Karan Khaturia was supported by University of Zurich Forschungskredit grant no. FK-19-080. Edoardo Persichetti was supported by the U.S. National Science Foundation grant CNS-1906360.}
}


\author{Nicolas Aragon             \and
        Marco Baldi                \and
        {Jean-Christophe} Deneuville \and
        Karan Khathuria            \and
        Edoardo Persichetti        \and
        Paolo Santini         
}


\institute{N. Aragon \at
              XLIM-MATHIS, University of Limoges, Limoges, France \\
              \email{nicolas.aragon@unilim.fr}           
           \and
           M. Baldi and P. Santini \at
              Department of Information Engineering, Marche Polytechnic University, Ancona, Italy \\
              \email{\{m.baldi,p.santini\}@staff.univpm.it}           
           \and
           {J.-C.} Deneuville \at
              Ecole Nationale de l'Aviation Civile, University of Toulouse, Toulouse, France \\
              \email{jean-christophe.deneuville@enac.fr}           
           \and
           K. Khathuria \at
              Institute of Mathematics, University of Zurich, Zurich, Switzerland \\
              \email{karan.khathuria@math.uzh.ch}           
           \and
           E. Persichetti \at
              Department of Mathematical Sciences, Florida Atlantic University, Boca Raton FL, USA \\
              \email{epersichetti@fau.edu}           
}

\date{Received: date / Accepted: date}

\maketitle

\begin{abstract}
We present an attack against a code-based signature scheme based on the Lyubashevsky protocol that was recently proposed by Song, Huang, Mu, Wu and Wang (SHMWW).
The private key in the SHMWW scheme contains columns coming in part from an identity matrix and in part from a random matrix.
The existence of two types of columns leads to a strong bias in the distribution of set bits in produced signatures.
Our attack exploits such a bias to recover the private key from a bunch of collected signatures.
We provide a theoretical analysis of the attack along with experimental evaluations, and we show that as few as 10 signatures are enough to be collected for successfully recovering the private key.
As for previous attempts of adapting Lyubashevsky’s protocol to the case of code-based cryptography, the SHMWW scheme is thus proved unable to provide acceptable security.
\highlight{This confirms that devising secure code-based signature schemes with efficiency comparable to that of other post-quantum solutions (e.g., based on lattices) is still a challenging task.}
\end{abstract}

\section{Introduction}
\label{sec:intro}

Digital signature schemes are a class of cryptographic primitives designed to provide a digital equivalent to their paper counterpart, namely to authenticate the original issuer of a document. Efficient constructions of signature schemes have been proposed alongside the advent of public key cryptography~\cite{RivShaAdl78}. Since then, a long line of research has aimed at making these constructions more efficient, by reducing the public key size and/or shortening the signature. While many well-established and widespread signature schemes rely on integer factorization, the most efficient constructions rely on the intractability of extracting discrete logarithms over the additive group of points on an elliptic curve. In 1994, assuming the existence of a sufficiently large quantum computer, Shor~\cite{FOCS:Shor94} presented an algorithm to solve both problems in polynomial time (as opposed to the best known classical algorithms, that require sub-exponential time). Finding quantum-safe alternatives to cryptosystems relying on the hardness of number theory problems is therefore of prime importance.\medskip

Among the quantum-safe alternatives, schemes based on Euclidean lattices and error-correcting codes stand as the most promising candidates. The latter defines the area known as code-based cryptography, which was initiated by McEliece~\cite{M78} in 1978, and essentially relies on the intractability of decoding random linear codes, a problem that has been proved to be NP-complete~\cite{BMvT78}. While it is relatively easy to build secure code-based public-key encryption schemes (for which the original McEliece approach is still robust), \highlight{obtaining efficient and secure digital signature schemes using the standard code-based approach (Hamming metric and syndrome decoding) is considerably more challenging.}
\medskip

Two methods are commonly used to design such schemes. The first one, the ``hash-and-sign'' paradigm that works very well for some traditional primitives (e.g. RSA), appears to be rather inadequate for code-based schemes. In fact, when relying on the hardness of decoding in the Hamming metric~\cite{BMvT78, barg1994some}, the difficulty of efficiently sampling decodable syndromes leads to protocols that are either inefficient or insecure (or both). 
\highlight{CFS~\cite{AC:CouFinSen01}, which historically dates as the first one in this category, is still technically unbroken (despite the introduction of a distinguisher~\cite{FGOPT10}) but fails to be practical due to its long signing times and large key sizes.}
\highlight{The latest hash-and-sign scheme, Wave~\cite{AC:DebSenTil19}, follows a new approach based on decoding of vectors of very large weight. In Wave, the public-key size grows quadratically in the security parameter, which is an important improvement over CFS. However, Wave still requires a public key of over 3 megabytes for 128 bits of classical security, and signing times of about 0.3 seconds.}
The second method, which consists of converting an identification scheme via Fiat-Shamir, typically results in very long signatures, due to the necessity of repeating the underlying Sigma protocol many times. The first code-based scheme of this type was proposed by Stern~\cite{SternZK} in '93, and the approach was successively refined through several subsequent works~\cite{Veron97, Cayrel2010, Aguilar2011, Bellini2019, less}. 
Yet, the signature sizes that one can obtain with this approach are still not optimal.\bigskip

A very promising solution, for lattice-based schemes, was given by Lyubashevsky in~\cite{EC:Lyubashevsky12}, leading to one of the top contenders for NIST's Post-Quantum standardization effort~\cite{nist}, Dilithium~\cite{NISTPQC-R2:CRYSTALS-DILITHIUM19}. The paradigm consists of a ``one-round'' application of an identification scheme \`a la Schnorr. This allows to obtain very compact signature sizes, as well as a simple and efficient signing procedure. 
As a consequence, there is a long history of works trying to adapt Lyubashevsky's protocol to the case of code-based cryptography. 
A first attempt was given by Persichetti~\cite{phdthesisEdoardo}, concluding that a simple conversion using both the traditional Hamming metric and the rank metric was unlikely to succeed. A subsequent work~\cite{onetime}, using quasi-cyclic codes and restricting to one-time usage, was susceptible to a similar attack~\cite{santini2019,DenGab20}. Finally, the authors in~\cite{durandal} present a solution based on the rank metric, including a slight modification of the Lyubashevsky protocol (with an additional masking error component), which appears to be secure and offers reasonable performance. 
However, there are still some doubts about information leakage in the scheme, and the security reduction leads to a rather convoluted, ad-hoc problem (named PSSI$^+$). Moreover, schemes based on the rank metric have shown vulnerabilities in recent times~\cite{bardet2020algebraic,AC:BBCGPSTV20}, which have undermined the community's confidence in this setting.
In the end, the problem of adapting the Lyubashevsky protocol through a decoding problem in the Hamming metric (which has been studied for decades and is now well-understood) is still open.

\paragraph{Contributions. }
In this paper we cryptanalyze the SHMWW scheme proposed in \cite{SHMWW}, which is another attempt at adapting the Lyubashevsky framework to coding theory.
The peculiarity of the SHMWW scheme consists in the structure of the private key, which is constructed according to an ad-hoc procedure that ensures the low weight of the signatures (this feature is at the core of the security proof).
However, the authors of \cite{SHMWW} have not considered that the distribution of set bits in the produced signatures is highly biased, according to the secret structure. 
This information leakage can be exploited to mount a full key-recovery attack, which can determine the private key after collecting a certain number of valid signatures. 
In light of our results, the SHMWW scheme can only be considered secure for one-time usage (at best); more generally, this work represents another evidence of the fact that
the Lyubashevsky framework appears to be not well-suited for coding theory.

\paragraph{Techniques. } 
Our proposed cryptanalysis of the SHMWW scheme can be divided into two steps. After having collected few signatures, one can perform a statistical test to distinguish between columns of weight one and the other columns in the private key.
This knowledge is then used to drive the information set choice in ISD algorithms: this way, the success probability for each ISD iteration becomes extremely high, and very few iterations are needed to recover each row of the private key. 
We first provide a theoretical analysis of a basic version of our attack, and show that it runs in time which is polynomial in the scheme parameters (this result, which comes with a closed formula for the running time of the attack, is summarized in Proposition \ref{prop:cost}).
Yet, this theoretical analysis is strongly conservative: as we show in Section~\ref{sec:expresults} with supporting experiments, the scheme can actually be broken with as little as 10 signatures (even 6 signatures are enough for attacking PARA-1 with a few days of running time). With as few as 32 signatures, the cryptanalysis successfully returns the secret key within 2 minutes for PARA-1 and 1 hour for PARA-2.

\paragraph{Related work. } The two independent works~\cite{cryptoeprint:2020:905} and~\cite{cryptoeprint:2020:923} described a similar strategy for efficiently attacking the SHMWW signature scheme.
Starting from those works, we present a unified cryptanalysis approach and an extended set of results. 

\section{Background and Notation}
\label{sec:background}

We start by introducing the notation used in this paper, which is kept as close as possible to that used in~\cite{SHMWW}.\\
We denote with $\mathbb{F}_q$ the finite field of $q$ elements. We use bold upper case (resp. lower case) letters to denote matrices (resp. vectors). The identity matrix of size $n\times n$ is denoted by $\6I_{n}$. Vectors are measured using the Hamming metric, and the Hamming weight of a vector $\6x$ is denoted by $\wt(\6x)$. The notation $\mathcal{V}_{n,w,q}$ indicates the set of all vectors of length $n$ and Hamming weight $w$, with components in $\mathbb{F}_q$. When the underlying field is clear from the context, this notation is simplified to $\mathcal{V}_{n,w}$.
We use $\mathfrak B(\rho)$ to denote the Bernoulli distribution with parameter $\rho$, and will write $x\sim\mathfrak B(\rho)$ to denote that $x$ is a random variable distributed according to $\mathfrak B(\rho)$.

\section{The SHMWW Signature Scheme}
\label{sec:shmww}
In this section we briefly recall the scheme in~\cite{SHMWW} and describe its main features. Public parameters are the integers $n,k,n',k',\ell,w_1,w_2,\dgv$, whose meaning will be clarified next. 
The scheme operates over the binary field, hence, for the remainder of this work, we will restrict our attention to the case $q=2$. The scheme also uses a ``weight restricted'' hash function $\WRH:\{0,1\}^*\to \mathcal{V}_{k',w_1}$, \highlight{\emph{i.e.}} a hash function that returns digests of length $k'$ and fixed weight $w_1$, which is not a novelty in code-based cryptography.
\medskip

Essentially, the authors propose a matricial version of the basic scheme described in~\cite[Table 7.17]{phdthesisEdoardo}, where the private key, instead of consisting of a single low-weight vector, is formed as a ``low-weight'' matrix, where by this we mean a matrix with a large number of zero entries. This is obtained by juxtaposing the systematic generator matrices $\mathbf{E}_1,\ldots,\mathbf{E_\ell}$ of $\ell$ distinct $[n',k']$ codes; the presence of the zeros is guaranteed by the identity matrix that appears as the leftmost block of a generator in systematic form. The matrix is then scrambled via both row and column permutations (the matrices $\mathbf{P}_1$ and $\mathbf{P}_2$, respectively) so that the final secret $\mathbf{E} = \mathbf{P}_1 \left[\mathbf{E}_1 | \cdots | \mathbf{E}_\ell\right] \mathbf{P}_2$ is essentially a large code (of length $n=n'\ell$) which should be, in the authors' intention, uncorrelated to the smaller codes forming it. The public key consists of a parity-check matrix $\mathbf{H}$ of a random $[n,k]$ code, and the matrix $\mathbf{S} = \mathbf{H}\mathbf{E}^\top$.\pagebreak

\begin{figure}[h]
\begin{algorithm}[H]
  \caption{KeyGen\label{alg:codekeygen1}}
  \begin{spacing}{1.3}
  \begin{algorithmic}[1]
    \Require{Public parameters $\textsf{params} = \left(n, k, n', k', \ell, w_1, w_2, \dgv\right)$.}
	\Ensure{$(\pk, \sk)$ with $\pk = \left(\mathbf{H}, \mathbf{S}\right)\in\F^{(n-k)\times n}\times\F^{(n-k)\times k'}$ and $\sk = \mathbf{E}\in\F^{k'\times n}$}
	\State
	Sample $\mathbf{H} \stackrel{\mathdollar}{\leftarrow} \F^{(n-k)\times n}$\State
	For $i=1,\dots,\ell$, sample $\mathbf{R}_i \stackrel{\mathdollar}{\leftarrow} \F^{k'\times (n'-k')}$ and set $\mathbf{E}_i\leftarrow\left(\mathbf{I}_{k'} | \mathbf{R}_i\right)$ \State
	Sample uniform random permutation matrices $\mathbf{P}_1, \mathbf{P}_2$ of respective sizes $k'\times k'$ and $n\times n$ \State
	Set $\mathbf{E} \leftarrow \mathbf{P}_1 \left[\mathbf{E}_1 | \cdots | \mathbf{E}_\ell\right] \mathbf{P}_2$ \State
	\Return{$\pk = \left(\mathbf{H}, \mathbf{S}= \mathbf{H}\mathbf{E}^\top\right), \sk = \mathbf{E}$}				
  \end{algorithmic}
  \end{spacing}
\end{algorithm}\vspace{-1.2cm}
\begin{algorithm}[H]
  \caption{Sign\label{alg:lyusign}}
    \begin{spacing}{1.3}
  \begin{algorithmic}[1]
    \Require{Public key $\pk$, private key $\sk$, and message $\mathbf{m} \in \left\lbrace0,1\right\rbrace^*$}
    \Ensure{Signature $\sigma=\left(\mathbf{z}, \mathbf{c}\right) \in \F^n\times\F^{k'}$ of message $\mathbf{m}$}
	\State
	Sample $\mathbf{e} \lar \mathcal{V}_{n,w_2}$ \State
	Compute $\6s\leftarrow\mathbf{H}\mathbf{e}^\top$ and $\mathbf{c} \leftarrow \WRH\left(\mathbf{m}\parallel \6s\right)$ 
	\State
	Set $\mathbf{z} \leftarrow \mathbf{c}\mathbf{E} + \mathbf{e}$\label{alg:codesign:sig}\State
	\Return{$\sigma=\left(\mathbf{z}, \mathbf{c}\right)$} \label{alg:codesign:proba1}
  \end{algorithmic}
    \end{spacing}
\end{algorithm}\vspace{-1.2cm}
\begin{algorithm}[H]
  \caption{Verify\label{alg:codeverif1}}
  \begin{spacing}{1.3}
  \begin{algorithmic}[1]
    \Require{Public key $\pk$, message $\mathbf{m}$, and signature $\sigma=\left(\mathbf{z}, \mathbf{c}\right)$}
	\Ensure{\textsf{Accept} if $\sigma$ is a valid signature of $\mathbf{m}$, \textsf{Reject} otherwise}
	\If{$\wt(\6z) \leq \ell\left(w_1 + n' - k'\right) + w_2$}
	\State
	Compute $\hat{\6s}\leftarrow\mathbf{H}\mathbf{z}^\top - \mathbf{S}\mathbf{c}^\top$
	\If{$\WRH\left(\mathbf{m}\parallel \hat{\6s}\right)=\6c$}\State
	\Return{\textsf{Accept}}
	\Else
	\State
	\Return{\textsf{Reject}}
	\EndIf
		\Else
	\State
	\Return{\textsf{Reject}}
	\EndIf
  \end{algorithmic}
    \end{spacing}
\end{algorithm}
\vspace{-0.95cm}
\caption{\label{fig:shmww}Song \etal code based proposal~\cite{SHMWW}.}
\end{figure}

\mbox{}
\vspace{-0.7cm}

To sign a message $\mathbf{m}$, a mask $\mathbf{e}$ of small weight $w_2$ is sampled uniformly at random, then committed by its syndrome, together with the message, to get the challenge $\mathbf{c} = \WRH\left(\mathbf{m} \parallel \mathbf{H}\mathbf{e}^\top\right)$. The response $\mathbf{z}$ to this challenge is the product of the private key and the challenge, hidden by the committed mask: $\mathbf{z} = \mathbf{c}\mathbf{E} + \mathbf{e}$. The signature $\sigma$ consists of the challenge and the response: $\sigma = \left(\mathbf{z}, \mathbf{c}\right)$. Note that no rejection sampling is performed during the signing process, unlike the original version of Lyubashevsky. Verification then proceeds accordingly with the dimensions of the objects in question, with the low ``weight'' of the secret matrix $\6E$ guaranteeing the low Hamming weight of the first component of the signature (the response vector $\6z$). The second component (the challenge vector $\6c$) is formed via the weight restricted hash function to ensure the final Hamming weight is below the desired threshold (parameters are chosen such that this is slightly above the GV bound). The algorithms comprising the SHMWW signature scheme are presented in detail in Fig.~\ref{fig:shmww}.

\paragraph{Parameter selection. } In~\cite{SHMWW}, the authors study the impact of applying Prange's Information Set Decoding (ISD) algorithm for both ``direct and indirect'' key-recovery attacks. This essentially provides parameters $n, k$, $\dgv$ and $w_2$; the other parameters follow by the Gilbert-Varshamov bound and by choosing a value for $\ell$:
\begin{equation}\label{eq:gvparams}
\highlight{\ell}\left(w_1 + n' - k'\right) + w_2 \leq \dgv.
\end{equation}  The proposed parameters are recalled in Table~\ref{tab:SHMWWparameters}.
\setlength{\tabcolsep}{5pt}
\begin{table}
\centering
\begin{tabular}{l*{11}{c}}
\toprule
Instance & $n$ & $k$ & $n-k$ & \highlight{$\ell$} & $n'$ & $k'$ & $n'-k'$ & $w_1$ & $w_2$ & $d=\dgv$ & $\lambda$\\
\midrule
Para-1 & 4096 & 539 & 3557 & 4 & 1024 & 890 & 134 & 31 & 531 & 1191 & 80\\
Para-2 & 8192 & 1065 & 7127 & 8 & 1024 & 880 & 144 & 53 & 807 & 2383 & 128\\
\bottomrule
\\
\end{tabular}
\caption{\label{tab:SHMWWparameters}Original SHMWW parameters~\cite{SHMWW} for $\lambda$ bits of security.}
\end{table}


\begin{table}
\centering
\begin{tabular}{lrrr}
\toprule
instance & keygen & sign & verif \\
\midrule
PARA-I & 415.98 & 3.81 & 4.48\\
PARA-II & 2,197.27 & 17.00 & 19.45 \\
\bottomrule \\
\end{tabular}
\caption{\label{tab:timingsSHMWW}Running times (ms) for the SHMWW signature scheme primitives. The timings were obtained by generating $10^3$ key generations, for each of which we generated $10^3$ signatures and verified them. Notice that the message signed was directly sampled as a vector of small weight $w_1$, instead of resorting to a weight restricted hash function as described in~\cite{SHMWW}.}
\end{table}

\section{Description of the attack}
\label{sec:attack}

\begin{figure}[ht]
\begin{subfigure}{\textwidth}
  \centering
  \begin{tikzpicture}

\draw[dashed] (-4.15,-0.54-0.4+0.13) rectangle (-2.2,1.13-0.4+0.13);

\draw[black, fill = mygray] (-4.15+2.1,-0.54-0.4+0.13) rectangle (-2.2+2.1, 1.13-0.4+0.13);


\draw[black, fill = mygray] (-4.15+4.35+2.1+0.1, -0.54-0.35+0.08) rectangle (-2.2+4.35+2.1+0.1, 1.13-0.35+0.08);


\draw [dashed] (-4.15+4.35, -0.54-0.35+0.08) rectangle (-2.2+4.35, 1.13-0.35+0.08);



\node at (0,0) {$\begin{bmatrix}\begin{matrix}\hspace{2mm}\begin{NiceMatrix}[columns-width = 0.41cm]1 & 0 & 0 & 0\\
0 & 1 & 0 & 0\\
0 & 0 & 1 & 0\\
0 & 0 & 0 & 1\end{NiceMatrix}
 & \begin{NiceMatrix}[columns-width = 0.41cm]
0 & 1 & 0 & 1\\
1 & 1 & 1 & 1\\
1 & 1 & 0 & 1\\
0 & 1 & 1 & 0
\end{NiceMatrix}\end{matrix} &
\begin{matrix}\begin{NiceMatrix}[columns-width = 0.41cm]1 & 0 & 0 & 0\\
0 & 1 & 0 & 0\\
0 & 0 & 1 & 0\\
0 & 0 & 0 & 1\end{NiceMatrix}
 & \begin{NiceMatrix}[columns-width = 0.41cm]
1 & 1 & 0 & 1\\
1 & 0 & 1 & 0\\
0 & 1 & 1 & 0\\
0 & 0 & 1 & 1
\end{NiceMatrix}\end{matrix}\hspace{2mm}\end{bmatrix}$};


\node at (-3.4,2) (T0) {Identity columns};
\draw[->] (T0) edge (-3.175,0.93);
\draw[->] (T0) edge (1.175,0.93);


\node at (3.7,2) (J) {Random columns};
\draw[->] (J) edge (-2.92+2.1,0.93);
\draw[->] (J) edge (-2.92+2.1+4.25,0.93);

\end{tikzpicture} 
  \caption{ }
\end{subfigure}
\begin{subfigure}{\textwidth}
  \centering
  \begin{tikzpicture}

\draw[black, fill = mygray] (-4.15,-0.81) rectangle (-3.25,0.86);

\draw[black, fill = mygray] (-2,-0.81) rectangle (-0.5,0.86);

\draw[black, fill = mygray] (1.75,-0.81) rectangle (2.85,0.86);

\draw[black, fill = mygray] (3.4,-0.81) rectangle (3.85,0.86);

\node at (0,0) {$\begin{bmatrix}\begin{matrix}\hspace{2mm}\begin{NiceMatrix}[columns-width = 0.41cm]
0 & 1 & 1 & 0\\
1 & 1 & 0 & 0\\
1 & 0 & 0 & 0\\
0 & 0 & 0 & 1
\end{NiceMatrix}
 & \begin{NiceMatrix}[columns-width = 0.41cm]
0 & 0 & 1 & 0\\
1 & 1 & 0 & 1\\
0 & 1 & 1 & 0\\
1 & 1 & 0 & 0
\end{NiceMatrix}\end{matrix} &
\begin{matrix}\begin{NiceMatrix}[columns-width = 0.41cm]
0 & 0 & 1 & 1\\
0 & 0 & 0 & 0\\
0 & 1 & 0 & 0\\
1 & 0 & 0 & 1
\end{NiceMatrix}
 & \begin{NiceMatrix}[columns-width = 0.41cm]
1 & 0 & 1 & 0\\
1 & 1 & 1 & 0\\
1 & 0 & 1 & 1\\
0 & 0 & 1 & 0
\end{NiceMatrix}\end{matrix}\hspace{2mm}\end{bmatrix}$};

\node at (0,2) (Ir) {$\mathcal I_R=\{1,2,5,6,7,12,13,15\}$};
\draw[->] (Ir) edge (-3.7,0.93);
\draw[->] (Ir) edge (-1.25,0.93);
\draw[->] (Ir) edge (2.3,0.93);
\draw[->] (Ir) edge (3.625,0.93);

\end{tikzpicture}  
  \caption{ }
\end{subfigure}
\caption{Example of separation of identity and random columns, for a private key with $n' = 8$, $k' = 4$ and $\ell = 2$. Figure (a) shows the matrix $[\6E_1|\6E_2]$, while Figure (b) displays the private key after application of the permutations $\6P_1$ and $\6P_2$. In this example, we have chosen $\6P_1$ equal to the identity and $\6P_2$ being the matrix corresponding to the permutation \highlight{$\{3, 8, 10, 4, 1, 15, 5, 13, 11, 14, 16, 9, 2, 7, 6, 12\}$}}.
\label{fig:secret_key}
\end{figure}
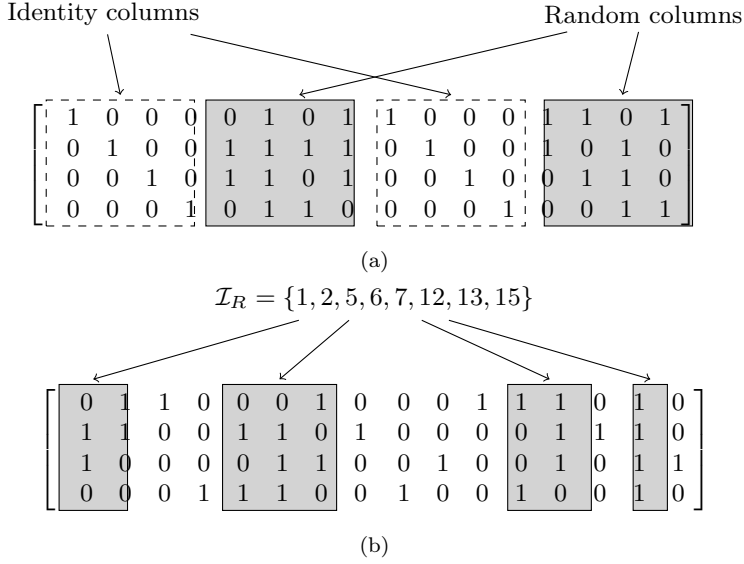

The columns of the private key $\6E$ in the SHMWW scheme can be divided into two groups, those due to identities, and those due to random submatrices: we will name the first ones as ``identity columns'', and the latter ones as ``random columns''. 
Finally, we will denote with $\mathcal I_R\subset \{1,\dots,n\}$ the set of integers pointing at random columns.
Let us represent the permutation defined by $\6P_2$ as $\{i_1,i_2,\cdots,i_n\}$, such that the $j$-th column is placed in position $i_j$; then, we have
$$\mathcal I_R = \{i_{k'+1},\dots,i_{n'},i_{n'+k'+1},\dots,i_{2(n')},\dots,i_{(\ell-1)n'+k'+1},\dots,i_{\ell n'}\}.$$
Note that the row permutation has no impact on the classification of the columns.  
For the sake of clarity, in Fig. \ref{fig:secret_key} we provide an example of this division for a toy private key where, for simplicity, we have chosen $\6P_1 = \6I_{k'}$). 

At a high level, our attack begins by recovering $\mathcal I_R$, \emph{i.e.} the location of random columns; then, exploiting this knowledge, we are able to recover each row of the secret $\6E$ using simple linear algebra.
In the next sections we formalize this procedure and provide a detailed analysis of its computational complexity.

\subsection{Leakage from the signatures}\label{sec:leakage}

The existence of two types of columns in the private key leads to a strong bias in the distribution of set bits in produced signatures, as we highlight in the following proposition.
\begin{proposition} \label{prop:zi_weight}
Let $\mathbf E$ be the private key and $\mathbf{z} = (z_1, \dots, z_n) = \mathbf{c} \mathbf{E} + \mathbf{e}$ be a signature. Further, let $\mathcal{I}_R$ be the set of random columns of $\mathbf{E}$. Then we have:

\begin{itemize}
\item $\rho_R = \mathrm{Pr}[z_i = 1]  = \frac{1}{2}$ if $i \in \mathcal{I}_R$;
\item $\rho_I = \mathrm{Pr}[z_i = 1] = \frac{w_1}{k'} + \frac{w_2}{n}(1 - 2 \frac{w_1}{k'})$ otherwise.
\end{itemize}
\end{proposition}

\proof
We know that $\mathbf{z} = \mathbf{c} \mathbf{E} + \mathbf{e}$, where $\mathbf{c}$ is a vector of length $k'$ and weight $w_1$ and $\mathbf{e}$ is a vector of length $n$ and weight $w_2$.
Since $w_1 \ll \frac{k'}{2}$, $\mathbf{c}$ has a much lower weight than a random vector of the same length. 

We first study the weight of each coordinate of the vector $\mathbf{z}' = \mathbf{c} \mathbf{E}$.
Let $z'_i$ be the $i$-th coordinate of $\mathbf{z}'$; there are two possibilities:
\begin{itemize}
\item if $i\in\mathcal I_R$, \emph{i.e.} if the $i$-th column of $\mathbf{E}$ is a random one, then $z'_i = 1$ with probability $\frac{1}{2}$;
\item if $i\not\in\mathcal I_R$, \emph{i.e.} if the $i$-th column of $\mathbf{E}$ is an identity one, then $z'_i = 1$ with probability $\frac{w_1}{k'}$. 
\end{itemize}
Now we want to compute the probability $\mathrm{Pr}[z_i = 1]$ that the $i$-th coordinate of $\mathbf{z}$ is of weight 1. Since $\mathbf{z}'$ and $\mathbf{e}$ are independent we have
\begin{align*}
\mathrm{Pr}[z_i = 1] &= \mathrm{Pr}[z'_i = 1] + \mathrm{Pr}[e_i = 1] - 2 \cdot\mathrm{Pr}[z_i = 1 \wedge e_i = 1]\\
&= \mathrm{Pr}[z'_i = 1] + \mathrm{Pr}[e_i = 1] \big(1 - 2\cdot\mathrm{Pr}[z'_i = 1]\big)
\end{align*}

Which gives the result by replacing $\mathrm{Pr}[z'_i = 1]$ by either $\frac{1}{2}$ or $\frac{w_1}{k'}$ depending on whether $i$ belongs to $\mathcal I_R$ or not, and $\mathrm{Pr}[e_i = 1]$ by $\frac{w_2}{n}$.

\qed

\begin{figure}[ht!]    
    \centering
    \includegraphics[keepaspectratio, width = \textwidth]{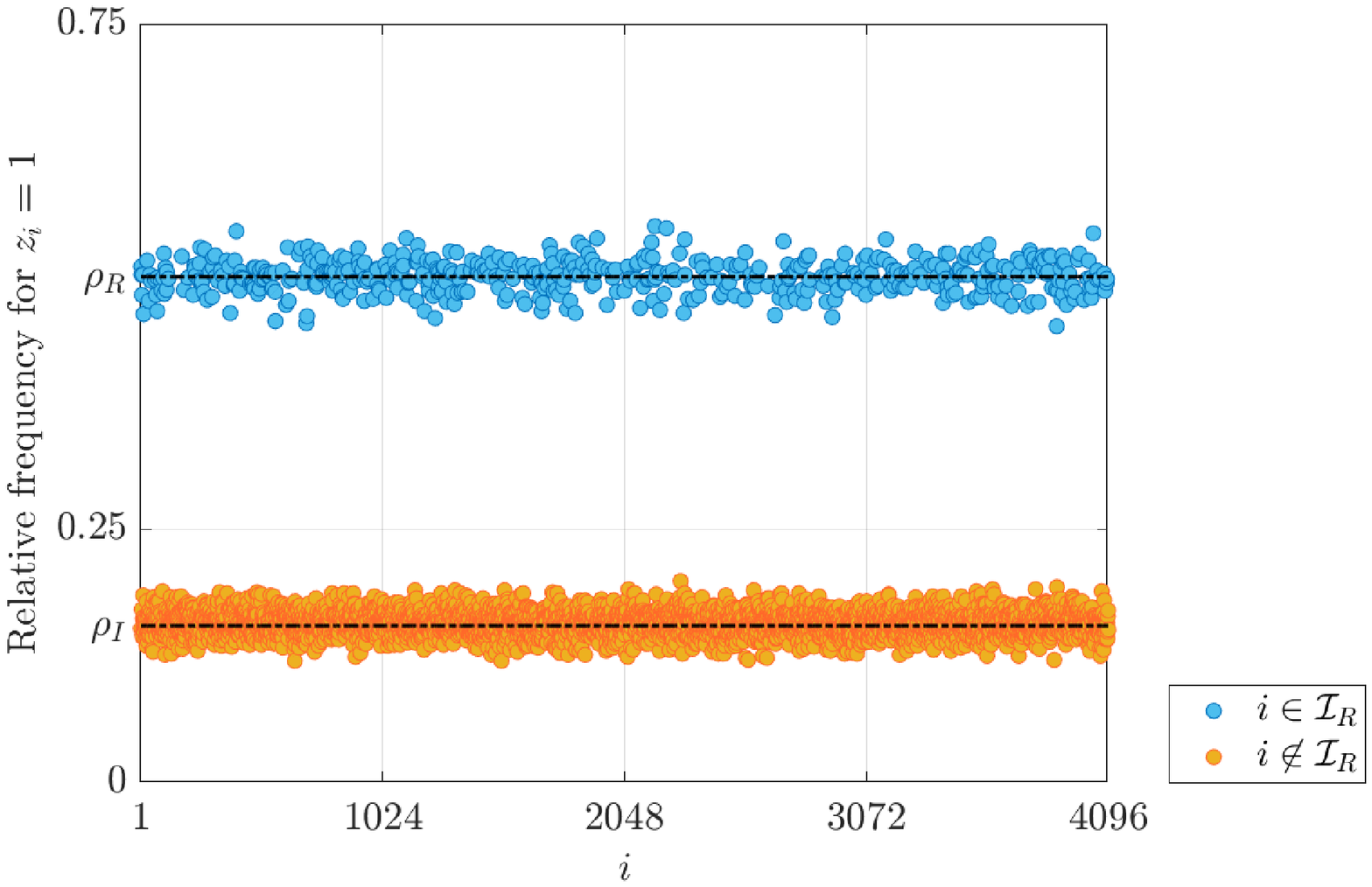}
    \caption{Relative frequency of $z_i = 1$ occurrences, for a random SHMWW Para-1 instance; for the experiment, we have considered a randomly generated private key and 1,000 signatures.    \label{fig:distinguishing}}
\end{figure}

\begin{table}
\centering

\begin{tabular}{|c|c|c|}
\hline
 & Para-1 & Para-2\\
\hline
$\rho_R$ & 0.5 & 0.5\\
\hline
$\rho_I$ & 0.155 & 0.147\\
\hline 
\end{tabular}

\caption{Values of $\mathrm{Pr}[z_i = 1]$ for the SHMWW parameter sets}
\label{tab:p_zi}
\end{table}
Table \ref{tab:p_zi} shows the values of $\mathrm{Pr}[z_i = 1]$ for the two SHMWW parameter sets that have been proposed in~\cite{SHMWW}.
As a consequence of Proposition \ref{prop:zi_weight}, we can distinguish between random and identity columns: when acquiring multiple signatures, the coordinates $z_i$ for which, on average, their weight is lower than $\frac{1}{2}$ are more likely to be the coordinates corresponding to columns of weight 1.
To provide an evidence of this fact, we have run numerical simulation on a random Para-1 instance; we have generated 1,000 signatures and, for each $i\in\{1,\dots,n\}$, we have computed the relative frequency with which the $i$-th entry is set. 
The obtained results are displayed in Fig. \ref{fig:distinguishing}.

In practice, one can guess $\mathcal I_R$ with a simple threshold criterion, which is applied after the observation of a bunch of honest signatures produced with the same key pair.
Let $N$ be the number of collected signatures, and denote with $(z_i)_j$ the $i$-th bit of the $j$-th collected one. 
For each $i\in\{1,\dots,n\}$, the adversary can compute  $\mu_i = \sum_{j=1}^N (z_i)_j$ and then apply the following rule
\begin{equation*}
\begin{split}
\mu_i \geq\delta N\implies \text{guess }i \in\mathcal I_R,\\
\mu_i <\delta N\implies \text{guess } i \not\in\mathcal I_R,
\end{split}
\end{equation*}
where $\delta\in \left(0 ; \frac{1}{2}\right)$.

A correct guess on $\mathcal I_R$ will be made if the values of $\mu_i$ are all $\geq \delta N$ for $i\in\mathcal I_R$, and are all lower than $\delta N$ for the remaining indexes.
We now derive the confidence level of this guessing phase, that is, the probability of making a correct guess for all indexes, as a function of the number of collected signatures $N$. 
To do this, we model each $\mu_i$ as the sum of $N$ independent random variables, following a Bernoulli distribution whose parameter depends on whether $i\in\mathcal I_R$ or not.
We recall Proposition \ref{prop:zi_weight} and, for a generic $i\in\mathcal I_R$, we estimate the probability of making a wrong guess as
\begin{align}
\label{eq:epsilon_R}
\epsilon_R \nonumber & = \mathrm{Pr}\left[\sum_{u=1}^Nx_u < \delta N\mid x_u\sim\mathfrak B\left(\rho_R = 1/2\right)\right]\\
& = 2^{-N}\cdot \sum_{u = 0}^{\left\lfloor \delta N\right\rfloor}\binom{N}{u}.
\end{align}
In an analogous way, in the case of $i\not\in\mathcal I_R$, we have that each $\mu_i$ is the sum of $N$ Bernoulli variables with parameter $\rho_I = \frac{w_1}{k'}+\frac{w_2}{n}\left(1-2\frac{w_1}{k'}\right)$; thus, we estimate the probability of wrongly guessing as 
\begin{align}
\label{eq:epsilon_I}
\epsilon_I \nonumber & = \mathrm{Pr}\left[\sum_{u=1}^Nx_u \geq \delta N\mid x_u\sim\mathfrak B\left(\rho_I\right)\right]\\
& = \sum_{u=\left\lceil \delta N\right\rceil}^N\binom{N}{u}\rho_I^u(1-\rho_u)^{N-u}.
\end{align}
Assuming that all values $\mu_i$ are independent, we have that the confidence level of the statistical test in the guessing phase, \emph{i.e.} the probability of correctly guessing all indexes, is
\begin{equation}
\label{eq:alpha}
\alpha = (1-\epsilon_R)^{|\mathcal I_R|} (1-\epsilon_I)^{n-|\mathcal I_R|}= (1-\epsilon_R)^{\ell(n'-k')}(1-\epsilon_I)^{\ell k'}.
\end{equation}
It is intuitively seen that, for an appropriate choice of $\delta$, the confidence level $\alpha$ rapidly grows with $N$; to further provide an understanding of this fact, we consider the following proposition.
\begin{proposition}\label{prop:N_star}
Let us assume that the values $\mu_i = \sum_{j = 1}^N(z_i)_j$ are independent and uncorrelated random variables.
Let $\alpha^* \in (0 ; 1)$,  $\delta\in \left(0 ; \frac{1}{2}\right)$ and
\begin{equation*}
\label{eq:N_star}
N^* = \max\left\{ \frac{4}{(1-2\delta)^2} \ln \left(\frac{2\ell (n'-k')}{1-\alpha^*}\right), \frac{(\delta+\rho_I)}{(\delta-\rho_I)^2} \ln\left(\frac{2\ell k'}{1-\alpha^*}\right) \right\}.\end{equation*}
Then, the confidence level of the test, \highlight{\emph{i.e.}} the probability of correctly guessing whether $i\in\mathcal I_R$ or not for all $i\in\{1,\cdots,n\}$, using $\delta$ as threshold and $N \geq N^*$ as the number of collected signatures, is not lower than $\alpha^*$.
\end{proposition}
The proof of the proposition, which makes use of the well known Chernoff bound, is provided in Appendix \ref{appendix:computing_N}.

\subsection{ISD complexity with the knowledge of positions of random columns}


Once $\mathcal{I}_R$ is known, we can recover $\mathbf{E}$ line by line by applying any Information Set Decoding (ISD) algorithm, such as Prange's algorithm \cite{prange}. 
\highlight{We briefly recall the definition of an information set and Prange's algorithm. An information set of an $[n,k]$ code is a subset $\mathcal I$ of $\{1,\ldots,n\}$ such that the columns of a parity-check matrix $\mathbf H$ indexed outside $\mathcal{I}$ form a non-singular matrix. Given as an input a parity-check matrix $\mathbf H$ and a syndrome $\mathbf s$, Prange's algorithm finds an error vector $\mathbf e$ of given weight $w$ such that $\mathbf H \mathbf e^\top = \mathbf s^\top$. The algorithm is based on the fact that if the support of the error vector $\mathbf e$ lies outside an information set, then the error vector can be recovered in polynomial time by solving a linear system of $n-k$ equations in $n-k$ variables.}

In order to recover the $j$-th line of $\mathbf E$, we apply Prange's algorithm on the parity-check matrix $\mathbf H$ with the $j$-th column of $\mathbf S$ as the syndrome. In addition, we choose an information set $\mathcal{I}$ such that $\mathcal{I}_R \subset \mathcal{I}$. This way we maximize the probability that every non-zero coordinates of the line we are trying to recover lies outside the information set.
\begin{proposition} \label{prop:ISD_prob}
The probability $p$ that the $\ell$ non-zero coordinates of $\mathbf{E}$ (the ones from the non-random columns) are included in $\mathcal{I}$ is:

\begin{equation} \label{eq:ISD_prob}
p = \frac{{n-k - (n'-k') \cdot \ell \choose \ell}}{{n - (n'-k') \cdot \ell \choose \ell}}.
\end{equation}

\end{proposition}

\proof

By choosing an information set $\mathcal{I}$ such that $\mathcal{I}_R \subset \mathcal{I}$, we have to choose $|\mathcal{I}| - |\mathcal{I}_R| = n - k - (n'-k') \cdot \ell$ columns at random and hope that the $\ell$ remaining non-null coordinates (from the identity matrices) are included in this set.

From this we deduce that the probability of success is the probability that the $\ell$ non-null coordinates that are distributed in $n - (n'-k') \cdot \ell$ positions are included in an information set of size $n - k - (n'-k') \cdot \ell$, hence the result.

\qed

We are now going to estimate the complexity of recovering the private key $\mathbf{E}$ given the knowledge of the set $\mathcal{I}_R$.

\begin{proposition}\label{prop:cost}
Given the knowledge of $\mathcal{I}_R$, recovering the private key $\mathbf{E}$ costs $\frac{k'(n-k)^3}{0.2887\cdot p}$ operations on average.
\end{proposition}

\proof

The complexity of solving a linear system to recover a line of $\mathbf{E}$ is $(n-k)^3$.
Since the SHMWW scheme only uses binary matrices, the probability that the matrix defining said linear system is invertible can be estimated as $\prod_{i = 1}^{n-k}1-2^{-i}\approx 0.2887$, and the probability $p$ that the system gives the correct solution is given by Proposition \ref{prop:ISD_prob}.
\\This has to be repeated for each of the $k'$ lines of $\mathbf{E}$, which gives the complexity in the thesis.
\qed

\subsection{Results}

Taking into account the results we have discussed in the previous section, we are now ready to present a complete attack on the scheme.
First, for the sake of completeness, in Fig. \ref{alg:key_recovery} we report the full procedure we use to attack the SHMWW scheme.
The work factor of an adversary attacking the scheme with this algorithm is estimated in the next proposition.
\begin{figure}
\centering

\fbox{
\begin{minipage}{.95\textwidth}
\textbf{Input:} $\mathbf{H}, \mathbf{S}$, a threshold value $\delta$, a set of signatures $(\sigma_1, \dots, \sigma_N) = ((\mathbf{z}_1, \mathbf{c}_1), \dots, (\mathbf{z}_N, \mathbf{c}_N))$

\textbf{Output:} the secret matrix $\mathbf{E}$

\begin{enumerate}
\item $\mathcal{I}_R = \emptyset$
\item For each $i$ from 1 to $n$:
\begin{itemize}
\item compute $\mu_i = \sum\limits_{j=1}^N (\mathbf{z}_j)_i$
\item if $\mu_i > N \cdot \delta$ then $\mathcal{I}_R = \mathcal{I}_R \cup \{i\}$
\end{itemize}
\item For each $i$ from 1 to $k'$:
\begin{itemize}
\item recover the $i$-th line of $\mathbf{E}$ by using an ISD algorithm and the knowledge of $\mathcal{I}_R$
\end{itemize}
\item Return $\mathbf{E}$
\end{enumerate}
\end{minipage}
}
\caption{Private key recovery of the SHMWW scheme}
\label{alg:key_recovery}
\end{figure}
\begin{proposition}
For each fixed $\alpha \in (0 ; 1)$ and $\delta \in (\rho_I ; 1/2)$, the algorithm described in Fig. \ref{alg:key_recovery} with a number of signatures equal to $N^*$ (as defined in Proposition \ref{prop:N_star}) returns the correct private key with probability at least $\alpha$, and has an average running time not greater than
$$n(N^*+1)+\frac{k'(n-k)^3}{0.2887\cdot p},$$
where $p$ is computed as in Proposition \ref{prop:ISD_prob}. 
\end{proposition}
\proof
In the first step (\emph{i.e.} instructions 1-2), the set $\mathcal I_R$ is guessed.
To do this, for each $j\in\{1,\dots,n\}$, one first computes $\mu_i$ (which costs $N^*$ operations), and then applies a threshold criterion, whose cost can be assumed to be equal to one elementary operation.
This justifies the first part of the complexity, while the second part simply corresponds to that of recovering the rows of $\6E$ through Prange's ISD (see Proposition \ref{prop:cost}).
For the success probability of the algorithm, we recall the analysis of Section \ref{sec:leakage}: to obtain a confidence level of $\alpha$, less than $N^*$ signatures are needed.
Then, using $N^*$ as the number of collected signatures allows us to derive a conservative estimate on the algorithm complexity.

\qed

\medskip

We are now able to assess the complexity of our attack on the proposed instances of the SHMWW scheme, targeting a confidence level of $\alpha = 0.9$:
\begin{enumerate}
    \item[-] for the Para-1 instance, designed for 80 bits of security, we choose $\delta = 0.3005$, yielding to $N^* = 250$; with these choices, our attack requires no more than $2^{48}$ operations;
    \item[-] for the Para-2 instance, designed for 128 bits of security, we choose $\delta = 0.3015$, yielding to $N^* = 264$; with these choices, our attack requires no more than $2^{52}$ operations.
\end{enumerate}

\subsection{Practical results and further considerations}

The results in the previous section, as captured by Proposition \ref{prop:ISD_prob}, already show that the SHMWW scheme can be broken in polynomial time, using a really limited number of signatures.
As we have already remarked, the analysis is rather conservative and, in a practical scenario, it is very likely that the attack can be performed with less significant effort; in this section, we motivate this claim with the aim of numerical results.
\medskip

First, the number of signatures the adversary needs to collect, to reach a desired confidence level, is significantly lower than that estimated as in Proposition \ref{prop:N_star}.
Indeed, the expression of $N^*$ is derived with the use of some conservative bounds, so this result is not surprising.
To support this claim, we have simulated the guessing phase, for the two originally proposed SHMWW parameters sets \cite{SHMWW}.
We have considered several values for the number $N$ of collected signatures and, for each value, we have simulated the guessing phase on 1,000 randomly generated key-pairs.
For each value of $N$, the value of $\delta$ has been chosen as the one maximizing the theoretical estimate of the confidence level expressed by \eqref{eq:alpha}.
The comparison between the theoretical estimates, and the actual confidence levels obtained through numerical simulations, is shown in Table \ref{tab:confidence_levels}. As we see, there is a very close correspondence between the theoretical values and the numerical ones: this fact constitutes a confirmation for the validity of our theoretical analysis.
Furthermore, it is easily seen that the number of signatures to reach a desired confidence levels are actually quite lower than those estimated through Proposition \ref{prop:N_star}.
Indeed, to reach $\alpha = 0.9$, we estimated $N^* = 250$ for Para-1 instances, and $N^* = 264$ for the Para-2 instances. 
As we see from the Table \ref{tab:confidence_levels}, such a confidence level can always be obtained after the collection of a much lower number of signatures.

\setlength{\tabcolsep}{5pt}
\begin{table}
\centering
\begin{tabular}{c|ccc|ccc}
\toprule
& \multicolumn{3}{c}{Para-1}&\multicolumn{3}{c}{Para-2}\\
$N$ & $\delta$ & Th. $\alpha$ & Emp. $\alpha$ & $\delta$ & Th. $\alpha$ & Emp. $\alpha$ \\
\midrule
10 & 0.300439 & $6.01\cdot 10^{-200}$ & 0 & 0.300872 & $1.22\cdot 10^{-383}$ & 0\\
30 & 0.333439 & $2.24\cdot 10^{-31}$ & 0 & 0.300872 & $1.20\cdot 10^{-51}$ & 0\\
50 & 0.320439 & $3.99\cdot 10^{-6}$ & 0 & 0.300872 & $2.39\cdot 10^{-9}$ & 0\\
70 & 0.314439 & $9.18\cdot 10^{-2}$ & 0.187 & 0.300872 & $2.73\cdot 10^{-2}$ & 0.076\\
90 & 0.311439 & 0.616 & 0.648 & 0.300872 & $0.508$ & 0.565\\
110 & 0.309439 & 0.903 & 0.923 & 0.300872 & $0.878$ & 0.9\\
130 & 0.308439 & 0.978 & 0.984 & 0.307872 & $0.976$ & 0.98\\
150 & 0.313439 & 0.996 & 1 & 0.306872 & $0.995$ & 0.998\\
170 & 0.312439 & 0.999 & 1 & 0.306872 & $0.999$ & 1\\
190 & 0.311439 & 0.999 & 1 & 0.305872 & $0.999$ & 1\\
\bottomrule
\end{tabular}
\caption{\label{tab:confidence_levels}
Confidence level of the guessing phase on the original SHMWW parameters~\cite{SHMWW}, for several values of the number $N$ of collected signatures. For each value of $N$, the value of $\delta$ has been chosen as the one maximizing the confidence level $\alpha$ expressed by \eqref{eq:alpha}. To numerically estimate the success rate, for each value of $N$, we have run the guessing phase on 1,000 randomly generated key-pairs.}
\end{table}
\medskip

We finally comment about the fact that, even when some additional indices are guessed inside $\mathcal I_R$, there is still some non null probability that an ISD algorithm can correctly return the rows of the private key.
In other words, if we choose threshold lower than the optimal one mentioned in Table \ref{tab:confidence_levels}, the statistical test fails (for some positions outside $\mathcal I_R$). In this way, we guess some additional indices in $\mathcal{I}_{R}$, but there is still some non null, and rather high, probability that an ISD algorithm can return the rows of the private key.
Thus, the scheme can still be attacked with a significantly lower number of collected signatures as described in the next section.

\section{Experimental results for the cryptanalysis of both parameter sets}
\label{sec:expresults}

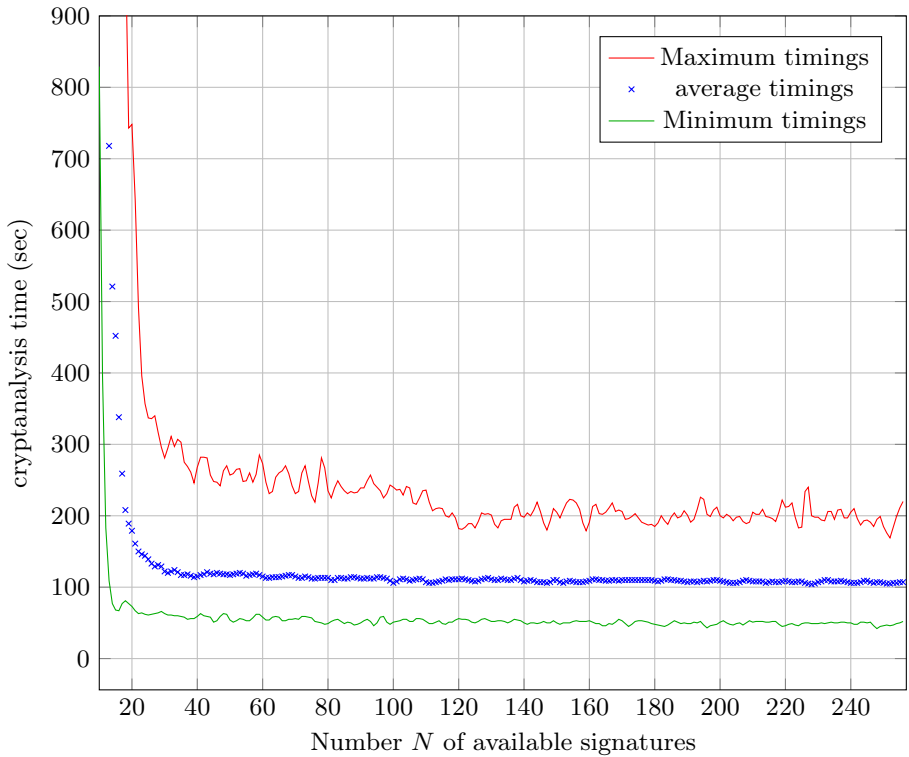
\begin{figure}[h]
\centering
\begin{tikzpicture}
	\begin{axis}[
		width=\textwidth,
		grid=major,
		xlabel={Number $N$ of available signatures},
		ylabel={cryptanalysis time (sec)},
		xmin=10,
		xmax=257,
		ymax=900,
		legend pos=north east,
	]
		
\addplot[red,name path=E]
file{plotdata/SfinalMaxes};
\addlegendentry{Maximum timings}

\addplot[blue,name path=C,mark=x,only marks,draw=none,mark options={scale=.75}]
file{plotdata/Sfinal};
\addlegendentry{average timings}

\addplot[green!67!black,name path=A]
file{plotdata/SfinalMins};
\addlegendentry{Minimum timings}

\end{axis}
\end{tikzpicture}

\caption{\label{fig:breakingPara1}Execution timings (sec) for breaking PARA-1 as a function of the number $N$ of signatures available to the adversary. Timings were averaged over a thousand executions.}
\end{figure}

To provide an evidence that the number of signatures required to successfully break the scheme is significantly lower than the theoretical value obtained in the previous section, we have run our cryptanalysis with different numbers of signatures available to the adversary for both parameter sets. For PARA-1, all cryptanalyses ran with 6 signatures or more were successful. This number had to be slightly increased for PARA-2 in order for the crytanalysis to complete within a week. As the number of available signatures increases, the execution timings quickly become very reasonable (minutes for PARA-1, hours for PARA-2). All the experiment results are reported on Fig.~\ref{fig:breakingPara1} for PARA-1 (targeting 80 bits of security) and Fig.~\ref{fig:breakingPara2} for PARA-2 (128 bits of security).

\begin{figure}[h]
\centering
\begin{tikzpicture}
	\begin{axis}[
		width=\textwidth,
		grid=major,
		xlabel={Number $N$ of available signatures},
		ylabel={cryptanalysis time (min)},
		xmin=15 ,
		xmax=257,
		ymin=0,
		ymax=540,
		legend pos=north east,
	]
		
\addplot[red,name path=E]
file{plotdata/SfinalMaxes2};
\addlegendentry{Maximum timings}


\addplot[blue,name path=C,mark=x,only marks,draw=none,mark options={scale=.75}]
file{plotdata/Sfinal2};
\addlegendentry{average timings}


\addplot[green!67!black,name path=A]
file{plotdata/SfinalMins2};
\addlegendentry{Minimum timings}

\end{axis}
\end{tikzpicture}

\caption{\label{fig:breakingPara2}Execution timings (min) for breaking PARA-2 as a function of the number $N$ of signatures available to the adversary. Timings were averaged over a hundred executions.}
\end{figure}

For these experiments, we use a threshold value obtained as a balanced combination of $\rho_R$ and $\rho_I$ in Prop.~\ref{prop:zi_weight}, where the weights correspond to the number of occurrences of each column type in $\6E$: 
\begin{equation}
    \label{eq:expthreshold}
    \delta = \left\lfloor\frac{\ell k'}{n}\cdot\left[\frac{w_1}{k'}+\frac{w_2}{n}\left(1-2\frac{w_1}{k'}\right)\right] + \frac{\ell\left(n'-k'\right)}{n}\frac{1}{2}\right\rfloor.
\end{equation}

We provide the resulting threshold for some numbers of collected signatures in Tab.~\ref{tab:expthreshold}.

\begin{table}[h]
    \centering
    \begin{tabular}{|c|c|c|c|c|c|c|c|c|c|c|}
        \cline{2-11}
        \multicolumn{1}{c}{} & \multicolumn{10}{|c|}{Number $N$ of available signatures} \\
        \cline{2-11}
        \multicolumn{1}{c|}{} & 10 & 16 & 24 & 32 & 64 & 128 & 160 & 192 & 224 & 256 \\
         \hline
         PARA-1 & 2 & 3 & 6 & 9 & 12 & 25 & 32 & 38 & 44 & 51 \\
        \cline{1-11}
         PARA-2 & 1 & 3 & 6 & 9 & 12 & 25 & 31 & 37 & 44 & 50 \\
         \hline
    \end{tabular}
    \caption{Experimental threshold values $N\cdot\delta$ to determine whether a column is random or not, according to Eq.~\eqref{eq:expthreshold}.}
    \label{tab:expthreshold}
\end{table}

Experiments were run over an Intel\textsuperscript{\textregistered{}} Xeon\textsuperscript{\textregistered{}} Gold 6230 CPU \@ 2.10GHz with Ubuntu 18.04, GCC 7.5.0 with compilation flags \texttt{-O3}, NTL 11.4.3, and gf2x 1.3.0. The reported execution timings have been averaged over 1000 executions. Both our implementation of the SHMWW scheme (without WRF) and the cryptanalysis are available at: \url{https://github.com/deneuville/cryptanalysisSHMWW_C}




\section{Conclusion}
\label{sec:conclusion}
We have presented an efficient cryptanalysis of the signature scheme recently proposed by Song \emph{et al.} in~\cite{SHMWW}, adapting Lyubashevsky's framework to coding theory. Our attack affects both parameter sets, and given its asymptotic complexity, discourages further parameter tweaks to patch this signature scheme. Our results are supported by a theoretical analysis and proof-of-concept implementations of the SHMWW signature scheme and its cryptanalysis. For both parameter sets, our attack requires as little as 10 signatures to fully recover the private key. Our results prove that the SHMWW signature scheme does not reach its claimed security, and should not be considered secure for more than one-time use.

\section*{Acknowledgement}

The authors thank Philippe Gaborit for insightful discussions on preliminary versions of this work.




%
%

\bibliographystyle{spbasic}      
\bibliography{sources_from_ADG/sign_bib/abbrev3, 
             sources_from_ADG/sign_bib/crypto_crossref,
              sources_from_ADG/sign_bib/sign_bib,
              sources_from_BKPS/References_modified,
              additional}


\appendix
\section{Computing the number of signatures for a desired confidence level} \label{appendix:computing_N}
We here prove Proposition \ref{prop:N_star}.
To bound the probabilities $\epsilon_R$ and $\epsilon_I$ which appear in \eqref{eq:alpha} we will use the Chernoff bound, which we recall in the following.
\begin{theorem}\textbf{Chernoff bound}\label{the_chernoff}
\\Let $X = \sum_{u = 1}^{M}x_u$, where the $x_u$ are all independent and $x_u\sim\mathfrak B( \rho)$; then
\begin{enumerate}
    \item[i)] $\mathrm{Pr}\left[X\leq (1 - \gamma) \rho M\right]\leq e^{-\frac{\gamma^2}{2}\rho M}$, for all $0<\gamma<1$;
        \item[ii)] $\mathrm{Pr}\left[X\geq (1+\gamma)\rho M\right]\leq e^{-\frac{\gamma^2}{2+\gamma}\rho M}$, for all $\gamma>0$.
\end{enumerate}
\end{theorem}
Applying condition i) of the Chernoff bound on \eqref{eq:epsilon_R}, we have $\rho = \frac{1}{2}$ and $\gamma = 1-2\delta$, such that
\begin{equation}
\label{eq:bound_epsilon_R}
    \epsilon_R\leq e^{-\frac{(1-2\delta)^2}{4}N} = \epsilon_R^*.
\end{equation}
In analogous way, applying condition ii) of the Chernoff bound on \eqref{eq:epsilon_I}, we have $\rho = \rho_I$ and $\gamma = \frac{\delta}{\rho_I}-1$, such that
\begin{equation}
\label{eq:bound_epsilon_I}
\epsilon_I\leq e^{-\frac{(\delta-\rho_I)^2}{\delta+\rho_I}N} = \epsilon_I^*.
\end{equation}
Using these bounds for $\epsilon_R$ and $\epsilon_I$, we derive the following inequality \textcolor{black}{on the success probability}
\begin{align*}
\label{eq:bound_alpha}
\alpha \geq (1-\epsilon_R^*)^{\ell(n'-k')}(1-\epsilon_I^*)^{\ell k'}.
\end{align*}
We first note that, regardless of the particular choice for $\delta$, the probabilities $\epsilon_R^*$ and $\epsilon_I^*$ decay exponentially with $N$; thus, we can always choose $N$ sufficiently high to make them extremely low.
Using a well known approximation, we have 
$$(1- \textcolor{black}{\epsilon_R^*}) ^{\ell (n'- k')}\approx 1 - \ell (n'-k') \textcolor{black}{\epsilon_R^*},$$
$$(1-\textcolor{black}{\epsilon_I^*}) ^{\ell k'}\approx 1 - \ell k' \textcolor{black}{\epsilon_I^*}.$$
\textcolor{black}{Now, let $$N \geq N^* = \max\left\{ \frac{4}{(1-2\delta)^2} \ln \left(\frac{2\ell (n'-k')}{1-\alpha^*}\right), \frac{(\delta+\rho_I)}{(\delta-\rho_I)^2} \ln\left(\frac{2\ell k'}{1-\alpha^*}\right) \right\}.$$ Then, $N \geq \frac{4}{(1-2\delta)^2} \ln \left(\frac{2\ell (n'-k')}{1-\alpha^*}\right)$ and \eqref{eq:bound_epsilon_R} implies that 
$$ \epsilon_R^* \leq \frac{1-\alpha^*}{2 \ell (n'-k')}, $$ and, $N \geq \frac{(\delta+\rho_I)}{(\delta-\rho_I)^2} \ln\left(\frac{2\ell k'}{1-\alpha^*}\right)$ and \eqref{eq:bound_epsilon_I} implies that $$\epsilon_I^* \leq \frac{1-\alpha^*}{2 \ell k'}.$$
Therefore, we obtain the following bound on the probability of success
\begin{align*}
    \alpha & \geq (1-\epsilon_R^*)^{\ell(n'-k')}(1-\epsilon_I^*)^{\ell k'} \\
    & \approx 1 - \ell (n'-k') \epsilon_R^* - \ell k' \epsilon_I^* + \ell^2 k' (n'-k')\epsilon_R^* \epsilon_I^*\\
& \geq  1 - \ell (n'-k') \epsilon_R^* - \ell k' \epsilon_I^* \\
& \geq \alpha^*.
\end{align*} }

\end{document}